# Aerosol Characteristics at a high-altitude station Nainital during the ISRO-GBP Land Campaign-II

**Auromeet Saha, P. Pant, U.C. Dumka, P. Hegde, Manoj K. Srivastava, and Ram Sagar**
Aryabhatta Research Institute for Observational Sciences (ARIES),
Manora Peak, Nainital-263 129, Uttaranchal.

## 1. Introduction:

During the second land campaign (LC-II) organised by ISRO-GBP, extensive ground-based measurements of aerosol characteristics were carried out over Manora Peak (29.4°N; 79.5°E; 1951 metres above mean sea level), Nainital (a high altitude station located in the Shivalik ranges of Central Himalayas) during the dry, winter season (December) of 2004. These measurements included the spectral aerosol optical depths (AOD), columnar water vapour content (W), Total Columnar Ozone (TCO), total number concentration ($N_T$) of near surface aerosols, mass concentration of black carbon ($M_B$), aerosol mass loading ($M_T$), and Global Solar Radiation. Based on these measured parameters, we present the results on the near-surface and columnar properties of atmospheric aerosols at Nainital.

## 2. Experimental details and Data:

The second land campaign (LC-II) was conducted during the period 1-31$^{st}$ December 2004. During this campaign, several instruments have been operated at Nainital for characterising the various aerosol properties. The location of the observation site (Nainital) is shown in the India map in Fig.1a and photograph of the sampling location is shown in Fig.1b.

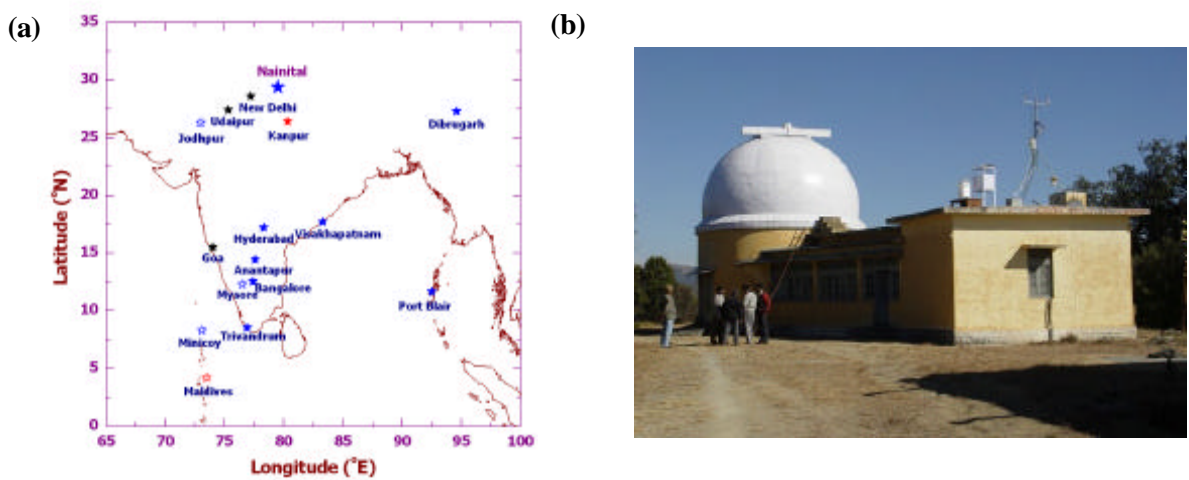

**Fig. 1.** (a) ISRO-GBP Network Stations, with the Observation Site (Nainital) as the northern-most station; (b) Photograph of the sampling location with the instruments kept on the roof-top of the building.

The instruments used in the present study consisted of (a) Multi-Wavelength solar Radiometer (MWR), measuring the AOD at ten wavelengths (centred at 380, 400, 450, 500, 600, 650, 750, 850, 935 and 1025 nm) and W, (b) Microtops-II Sun Photometer (Solar Light Co.) measuring the instantaneous AODs at six different wavelengths (centred at 380, 440, 500, 675, 870, and 1020 nm) and W, (c) Microtops-II Ozonometer (Solar Light Co.), measuring the instantaneous values of Total Columnar Ozone; (d) Aerosol Spectrometer (GRIMM) measuring the number concentration and size distribution of near-surface aerosols;

(e) dual-channel Aethalometer (Magee Scientific) measuring the near-surface mass concentration of Black Carbon aerosols ($M_B$), (f) High Volume air Sampler (HVS) measuring the mass loading of near-surface aerosols ($M_T$), (g) Pyranometer (Campbell Scientific) for measuring the Global Solar Radiation.

The MWR, Microtops-II Sun Photometer and Ozonometer were operated regularly on all clear/partly-clear sky days. The raw data obtained from the MWR were analysed using the conventional Langley technique to deduce the columnar AOD ($\tau_{p\lambda}$). Usually, the Langley plot showed two different slopes on the same day, one for the forenoon (FN) part and the other for the afternoon (AN), and these data were analysed separately by considering the FN and AN data as two independent sets and the AODs for each set were estimated. More details on the instrumentation, data analysis and error budget are given elsewhere (*Moorthy et al.*, 1999; *Sagar et al.*, 2004). A total of 33 data sets were obtained from the MWR measurement during the study period. As Microtops-II gives the instantaneous values of AOD, W, and TCO based on its internal calibration, more data sets (as compared to MWR) was usually obtained with it. In the present study, more than 650 data sets were obtained (out of the 26 clear sky days) from the Microtops-II Sun Photometer and Ozonometer measurements.

Measurements of the number concentration and size distribution of near-surface aerosols were carried out using a GRIMM Aerosol Spectrometer (Model 1.108). It uses the light-scattering technology for single particle counts in which a semiconductor laser serves as the light source. The scattered signal from the particle passing through the laser beam is collected at approximately 90° by a mirror and transferred to a recipient diode. After a corresponding reinforcement, the signal of the diode passes through a multi-channel size classifier. A pulse-height analyser then classifies the signal transmitted in each channel. The Aerosol Spectrometer can be used for the continuous and near real-time measurement of aerosols in one of two basic modes: (a) particle counts as counts/litre or (b) mass as µg/m$^3$ at 15 different size distribution channels with the size range covering 0.3 to 20 µm. In the present study, the instrument was run in the count mode to obtain the aerosol number concentration, and the measurement cycle (time base) was set to 5 minutes. Continuous measurements of aerosol number concentration were carried out at Nainital and the data was available on all the 31 days in the month of December 2004.

Near real-time measurements of the mass concentration of Black Carbon (BC) aerosols were carried out using the dual-channel Aethalometer (Model AE-21), since 9$^{th}$ December 2004. The Aethalometer is a fully automatic instrument that uses a continuous filtration of ambient aerosols and optical transmission measurement technique to estimate the mass concentration of BC. The instrument aspirates ambient air using its inlet tube and its pump. The BC mass concentration is estimated by measuring the change in the transmittance of a quartz filter tape, on to which the particles impinge. The principal of operation and the other details are available elsewhere [*Babu and Moorthy*, 2002; *Hansen*, 2003]. In our present study, the Aethalometer was configured for a constant flow rate of 3.3 litre min$^{-1}$ and the time base was set to 5 minutes.

Bulk samples of aerosols were collected using a High Volume air Sampler (HVS), provided by the PRL group. During the study period, a total of 20 samples were collected, with 16 samples collected during the day-time and 4 samples during the night-time. From HVS measurements, the mass loading of the total suspended aerosols were estimated.

The Global Solar Radiation was measured using a Pyronometer. The measurement cycle was set at 20 minutes. From these measurements, the diurnal variation of solar radiation was studied.

## 3. Prevailing Meteorology:

The prevailing meteorology at Nainital during December comprises of a synoptic north-westerly circulation, dry ambient with low RH (~50 to 70%) and scanty rainfall. This season is also characterized by generally clear and cloud free sky conditions and absence of any major weather phenomena. However, during the present study, it was observed that during most of the days, the morning skies were in general clear (cloud-free), followed by cloudy and sometimes overcast sky in the afternoon.

## 4. Results on the Columnar Properties of Aerosols

### 4.1. Aerosol Optical Depth (AOD):

In the present study, since the AOD measurements were obtained from two different instruments (MWR and Microtops-II), and some wavelengths were common to both, it was possible to inter-compare the results obtained from both these instruments and thereby ascertain their mutual consistency, reliability and inter-usability. Fig.2a shows the inter-comparison of the monthly-mean spectral AODs derived from MWR and Microtops-II. The results show that both the instruments produced quite similar results and they remained stable and statistically consistent during the study period. In light of this, the results from either of these instruments are used independently, and complementarily, whenever and wherever required.

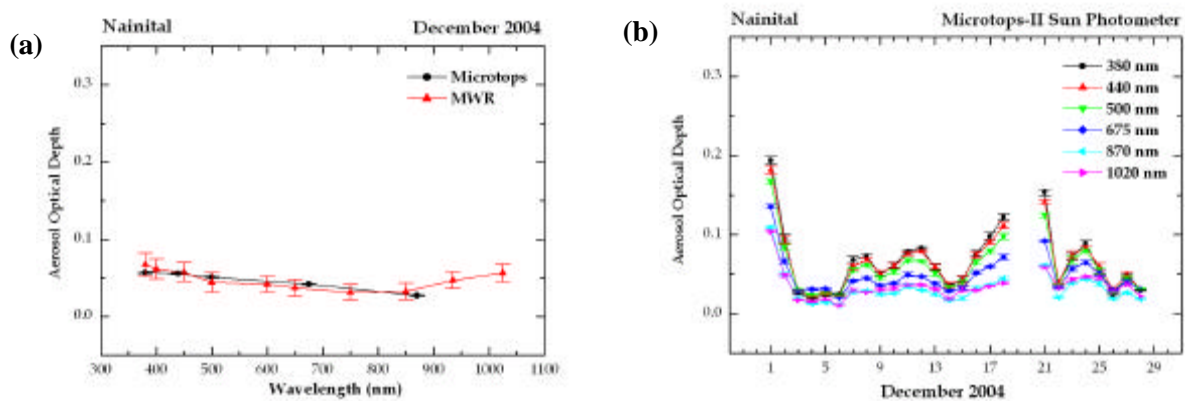

**Fig. 2. (a) Comparison of monthly-mean spectral AODs obtained from MWR and Microtops-II; (b) Temporal variation of AODs at all the wavelengths of Microtops-II.**

Fig.2b shows the temporal variation of AODs obtained from the Microtops-II Sun Photometer at all the six wavelengths. It can be clearly seen from these figures that the AODs (at all the wavelengths) shows significant day-to-day variability. The temporal variation of AODs shows moderately high values (>0.1 at 500nm) during $1^{st}$, $18^{th}$ and $21^{st}$ December and low values (~0.05) during rest of the period. The monthly-mean AOD at 500 nm was 0.056 (±0.037), which is typical for a remote, high altitude location.

### 4.2. Columnar Water Vapour Content (W):

Fig.3 shows the temporal variation of columnar water vapour content (W) derived from the Microtops-II measurements. The temporal variations clearly show the day-to-day variability, with highest value (~0.55 cm) observed on $1^{st}$ December and lowest value (~0.04 cm) on $26^{th}$ December. The monthly-mean value of W was 0.28 (±0.11) cm, which shows the prevalence of very dry environment at Nainital. Further, the figures (Fig.2b & Fig.3) also reveals that the variations of W are somewhat similar to that of AOD, thereby indicating that the columnar water vapour may also be contributing to the observed AODs, as several studies (for example, *Nair and Moorthy*, 1998) have shown that, an increase in RH and/or W causes a non-linear increase in AOD at all wavelengths.

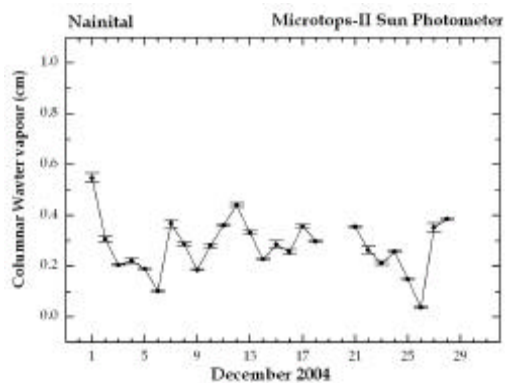

**Fig.3. Temporal variation of W.**

### 4.3. Total Columnar Ozone:

The temporal variation of total columnar ozone (TCO) obtained from the Microtops-II Ozonometer is shown in Fig.4a. Also shown in the figure are the temporal variations of Total Ozone retrieved from the Earth-Probe Total Ozone Mapping Spectrometer (TOMS) satellite. The results show that both the ground-based (Microtops-II) and satellite measurements (TOMS) show the similar trend in the ozone variations. The correlation between both the measurements (shown in Fig.4b) is very significant (with correlation coefficient of 0.96). However, TOMS under-estimates the total ozone values, by as much as 20 DU. The temporal variations of TCO (Fig.4a) shows that the Ozone concentrations were around ~260 DU during most of the days. However the TCO increased to high values during 21-26$^{th}$ December, during which values >300 DU were encountered. The monthly mean TCO was found to be 268 (±22) DU.

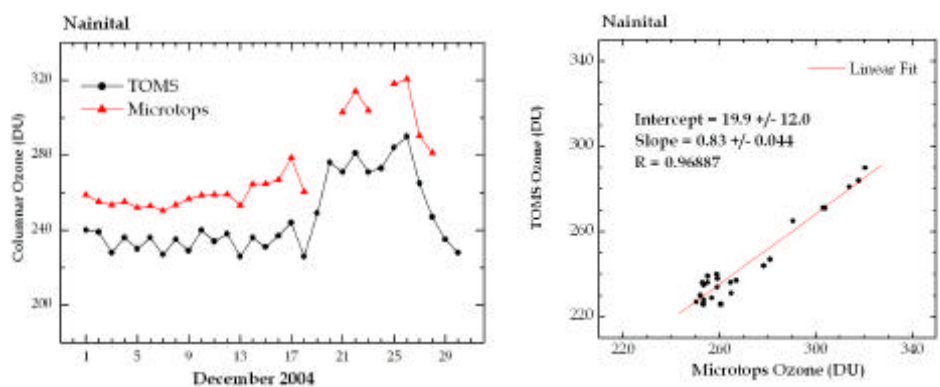

**Fig.4.  (a) Temporal variation of Ozone obtained from Microtops-II and TOMS; (b) Inter-comparison of Ozone obtained from Microtops-II and TOMS.**

### 5.   Results on the Near-Surface Properties of Aerosols

### 5.1. Aerosol Number Concentration:

Fig.5a shows the temporal variations of total number concentration ($N_T$) of near-surface aerosols, which shows significant day-to-day variability. Besides the day-to-day variability, $N_T$ also shows a well-defined diurnal variation (Fig.5b) with extremely low values occurring during night and early morning, which gradually increases after sunrise to attain a peak in the afternoon (between 14:00 to 16:00 hrs local time). The concentration thereafter decreases gradually at sunset and reaches to a very low value by night. The afternoon peak can be attributed to the vertical transport of aerosols from the near-by polluted urban and valley regions, which were initially confined to lower heights in the night and early morning due to the low-level inversions, but are released to greater heights as the boundary layer evolves.

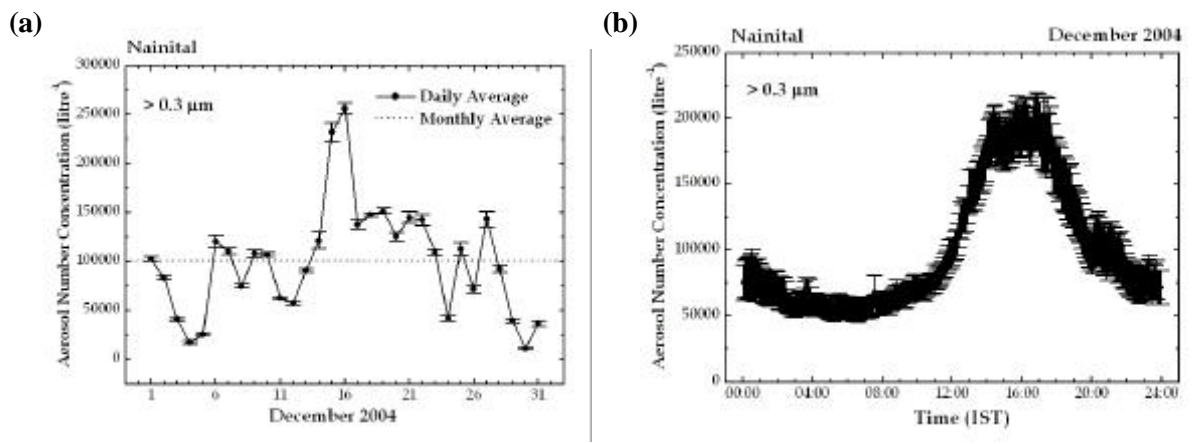

**Fig.5.** (a) Temporal variation of total number (>0.3 mm) concentration ($N_T$) of near-surface aerosols; (b) Monthly-mean diurnal Variation of $N_T$.

### 5.2. Black Carbon Mass Concentration:

Fig.6a shows the temporal variations of BC mass concentration and Fig.6b shows the month-mean diurnal variation. The temporal and diurnal variation of BC mass concentration shows similar variations as that $N_T$ (Fig.4), thereby showing similar processes (such as boundary layer dynamics) to be responsible in causing these variations. The monthly mean BC concentration was found to be 1.36 (±0.99) µg m$^{-3}$, which is very much lower compared to other urban locations such as Trivandrum, Goa, Hyderabad (*Babu and Moorthy*, 2002; *Lata and Badrinath*, 2003; *Leon et al.*, 2001).

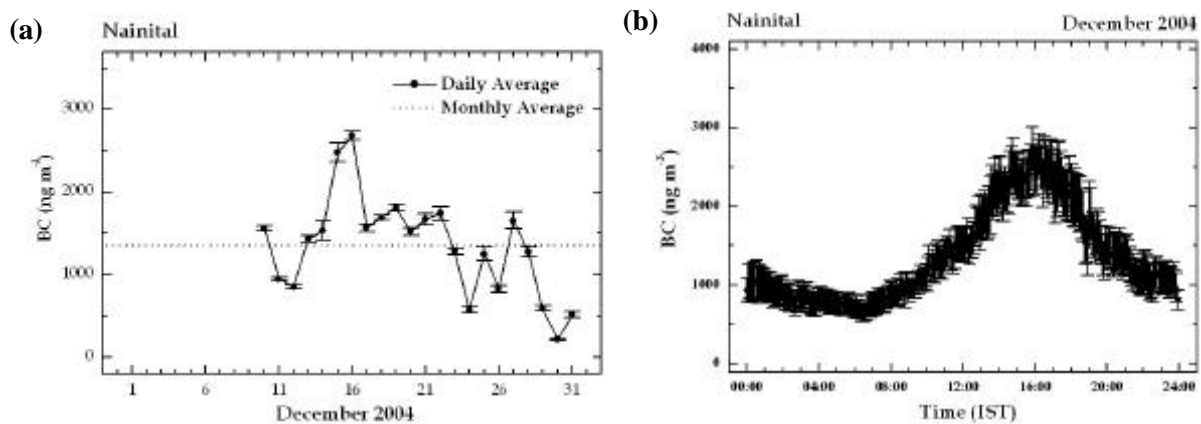

**Fig.6.** (a) Temporal variation of near-surface Mass Concentration ($M_B$) of BC; (b) Monthly-mean diurnal variation of $M_{BC}$.

### 5.3. Aerosol Mass Loading:

The mass loading ($M_T$) of total suspended aerosols particulates (TSP) was measured using a High Volume air Sampler. Since, the aerosol concentrations at Nainital were very low, the sampler was operated continuously for two-days during the daytime. For obtaining the nighttime concentrations, the sampling was done from the evening of first day till morning of the second day. Fig.7a shows the temporal variation of aerosol mass loading for the month of December 2004. The mass loading of TSP is in the range 20-40 µg m$^{-3}$, with a mean value of 27.1 (±8.3) µg m$^{-3}$. Further, it can also be seen from the Fig.6, that during most of the cases, the night-time concentrations are much lower than the day-time values, which can be attributed to the boundary layer dynamics.

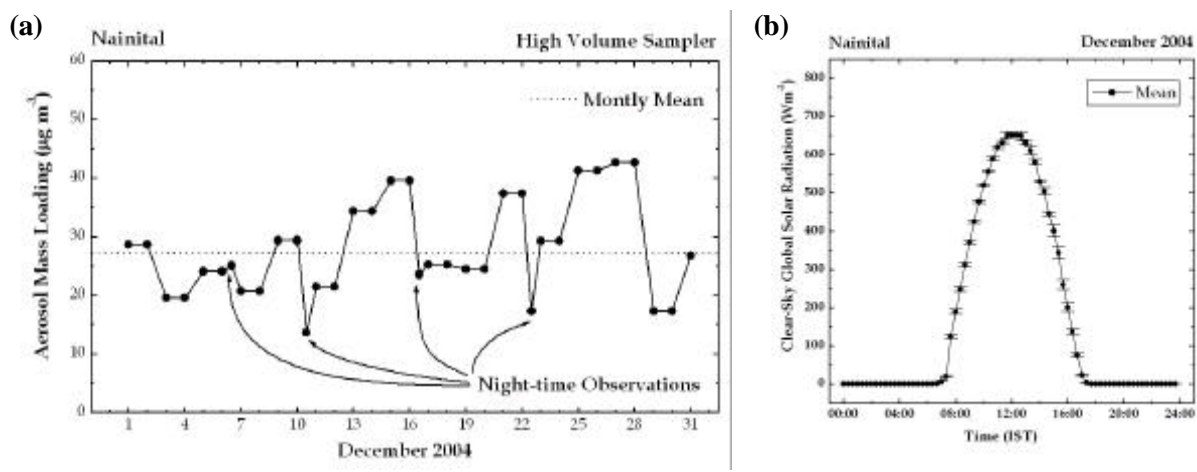

**Fig.7.** (a) Temporal variation of near-surface aerosol mass loading ($M_T$); (b) Diurnal variation monthly mean clear-sky Global Solar Radiation

### 5.4. BC Mass Fraction:

Even though BC contributes only a few percent to the total aerosol mass, it produces significant radiative effects. The apportionment of BC is thus very important in modelling the aerosol radiative properties. To estimate the BC mass fraction, we have used the monthly mean aerosol mass loading ($M_T$) and BC mass concentration ($M_B$). The share of BC to total aerosol mass at Nainital was found to be ~5%. It is very much lower than the values reported for the west coast location Trivandrum, which showed 12% share of BC during December (*Babu and Moorthy*, 2002).

### 5.6. Global Solar Radiation:

Fig.7b shows the diurnal variation of monthly mean global solar radiation at Nainital. The maximum mean was found to be ~650 W m$^{-2}$.

**Acknowledgements:**
This work was carried out as a part of Land-Campaign-II, under the ISRO-Geosphere Biosphere Program. The authors would like to express their sincere gratitude to Prof. M.M Sarin of Physical Research Laboratory, Ahmadabad for providing the High Volume Sampler.